


%
%

\hsize=14.cm
\vsize=20.cm   
\hoffset=1.2cm
\voffset=1.6cm
\baselineskip=13pt plus 1pt minus 1pt
\tolerance=1000
\parskip=0pt
\parindent=15pt

\def\normal{\baselineskip=13pt plus 1pt minus 1pt}

\nopagenumbers

\font\tenss=cmss10

\font\eightrm=cmr8
\font\ninerm=cmr9

\font\nineit=cmti9
\font\eightit=cmti8

\font\eightsl=cmsl8
\font\ninesl=cmsl9

\font\ninebf=cmbx9

\def\myref#1{[{#1}]}

\def\newsection#1{\vskip .6cm
          \leftline{\bf #1}  \vglue 5pt}

\def\hspace{~\hskip 1cm ~}

\def\next{~~~,~~~}
\def\big{\displaystyle \strut }

\def\N{\kappa}

\def\vn{ {\vec n} }
\def\vk{ {\vec k} }
\def\vl{ {\vec l} }
\def\vf{ {\vec f} }
\def\vg{ {\vec g} }
\def\vh{ {\vec h} }
\def\vp{ {\vec p} }
\def\vv{ {\vec v} }

\def\vtheta{ {\vec \theta} }
\def\vphi{ {\vec \phi} }

\def\vgamma{ {\vec \gamma} }
\def\vone{ {\vec 1} }

\def\btheta{ {\bar \theta} }

\def\bgamma{ {\bar \gamma} }
\def\bU{ U \kern-.75em\raise.65em\hbox{$-$} }
\def\bW{ W \kern-.9em\raise.65em\hbox{$-$} }
\def\bUq{ U \kern-.75em\raise.65em\hbox{$-${\eightit q}} }
\def\bUqj{ U_j \kern-1.05em\raise.65em\hbox{$-\,${\eightit q}} }
\def\bUqone{ U_1 \kern-1.05em\raise.65em\hbox{$-\,${\eightit q}} }
\def\bUqtwo{ U_2 \kern-1.05em\raise.65em\hbox{$-\,${\eightit q}} }

\def\x{ {\bf x} }
\def\y{ {\bf y} }

\def\zz{ \hbox{\tenss Z} \kern-.4em \hbox{\tenss Z} }

\def\L{ {\cal L} }
\def\tild{ {\tilde d} }
\def\ep{\epsilon}
\def\eps{\varepsilon^{\mu\nu\rho}}
\def\d{\partial}
\def\epdel{\,\varepsilon\partial\,}

\def\d{\partial}
\def\la{\raise.16ex\hbox{$\langle$} \, }
\def\ra{\, \raise.16ex\hbox{$\rangle$} }
\def\st{\, \raise.16ex\hbox{$|$} \, }
\def\go{\rightarrow}

\def\CS{ {\rm CS} }
\def\eff{ {\rm eff} }
\def\tot{ {\rm tot} }

\def\Dbar{ D \kern-.7em\raise.65em\hbox{$-$} }
\def\Dsquare{ D_k \kern-1.2em\raise.63em\hbox{$-{\scriptstyle 2}$} }
\def\Dbarky{ D_k \kern-1.2em\raise.63em\hbox{$-{\scriptstyle y}$} }
\def\DbarjI{ D_j \kern-1.2em\raise.63em\hbox{$-{\scriptstyle I}$} }
\def\DbarkI{ D_k \kern-1.2em\raise.63em\hbox{$-{\scriptstyle I}$} }

\def\psibar{ \psi \kern-.65em\raise.6em\hbox{$-$} }
\def\Lbar{ {\cal L} \kern-.65em\raise.6em\hbox{$-$} }


\global\newcount\refno
\global\refno=1 \newwrite\reffile
\newwrite\refmac
\newlinechar=`\^^J
\def \ref#1#2{\the\refno\nref#1{#2}}
\def\nref#1#2{\xdef#1{\the\refno}%
	\ifnum\refno=1\immediate\openout\reffile=refs.tmp\fi%
	\immediate\write\reffile{\noexpand\item{[\noexpand#1]}#2\noexpand\nobreak}
	\immediate\write\refmac{\def\noexpand#1{\the\refno}}%
	\global\advance\refno by1}


\def\semi{\hfil\noexpand\break ^^J}  

\def\refn#1#2{\nref#1{#2}}


\def\ap#1#2#3{{\nineit Ann.\ Phys.\ (N.Y.)} {\ninebf {#1}}, #3 (19{#2})}

\def\ijmpA#1#2#3{{\nineit Int.\ J.\ Mod.\ Phys.} {\ninebf {A#1}}, #3 (19{#2})}
\def\ijmpB#1#2#3{{\nineit Int.\ J.\ Mod.\ Phys.} {\ninebf {B#1}}, #3 (19{#2})}

\def\mplA#1#2#3{{\nineit Mod.\ Phys.\ Lett.} {\ninebf A{#1}}, #3 (19{#2})}

\def\plB#1#2#3{{\nineit Phys.\ Lett.} {\ninebf {#1}B}, #3 (19{#2})}

\def\np#1#2#3{{\nineit Nucl.\ Phys.} {\ninebf B{#1}}, #3 (19{#2})}
\def\prl#1#2#3{{\nineit Phys.\ Rev.\ Lett.} {\ninebf #1}, #3 (19{#2})}
\def\prB#1#2#3{{\nineit Phys.\ Rev.} {\ninebf B{#1}}, #3 (19{#2})}

\def\prp#1#2#3{{\nineit Phys.\ Report} {\ninebf {#1}C}, #3 (19{#2})}
\def\rmp#1#2#3{{\nineit Rev.\ Mod.\ Phys.} {\ninebf {#1}}, #3 (19{#2})}

\def\cline{\hfil\noexpand\break  ^^J}

\immediate\openout\refmac=refno.tex

\refn\Reviews{
 For reviews, see eg.
\semi F. Wilczek, {\nineit Fractional Statistics and Anyon Superconductivity},
(World Scientific 1990);
\semi E. Fradkin, {\nineit Field Theories of Condensed Matter Systems},
(Addison-Wesley 1991);
\semi A.L. Fetter, C.B. Hanna, and R.B. Laughlin, \ijmpB {5} {91} {2751};
\semi A. Zee,  {\nineit From Semionics to Topological Fluids},
ITP preprint NSF-ITP-91-129, (1991);
\semi D. Lykken, J. Sonnenschein,  and N. Weiss, \ijmpA {6} {91} {5155};
\semi S. Forte, \rmp {64} {92} {193};
\semi R. Iengo and K. Lechner,  \prp {213} {92} {179};
\semi D. Boyanovsky,   \ijmpA {7} {92} {5917};
\semi Y. Hosotani, {\nineit Neutral and Charged Anyon Fluids},
UMN-TH-1106/92 (To appear in {\nineit Int. J. Mod. Phys. {\ninebf B}}).      }

\refn\WenZeeOne{
X.G. Wen and A. Zee, {\nineit Nucl. Phys.} {\ninebf B15} (Proc. Suppl.) 135
(1990);
\prB {44} {91} {274};
{\ninebf B46}, 2290 (1992); \prl {69}{92} {953};
\semi J. Fr\" ohlich and A. Zee, \np {364} {91} {517}.   }

\refn\ZeeReview{
For reviews, see eg.
\semi A. Zee, in Ref. [1]; {\nineit Long Distance Physics of Topological
Fluids}, in the Proceedings of {\ninesl Low Dimensional Field Theories and
Condensed Matter Physics}, Kyoto, 1991; \ijmpB {5} {91} {1629}.   }

\refn\LeeRead{
D.H. Lee and M.P.A. Fisher, \prl {63} {89} {903};
\semi N. Read, \prl {65} {90} {1502}.   }

\refn\BlokWen{
B. Blok and X.G. Wen, {\nineit Phys. Rev.} {\ninebf B42}, 8133, 8145 (1990);
\prB {43} {91} {8337}.   }

\refn\WenZeeTwo{
X.G. Wen and A. Zee, \prl {69} {92} {1811}; {\nineit Tunneling
in Double Layered Quantum Hall Systems}, NSF-ITP-92-78 (1992).   }

\refn\GWW{
M. Greiter, X.G. Wen and F. Wilczek, \prl {66} {91} {3205};
\np {374} {92} {567};  {\nineit Paired Hall States in Double Layered Electron
Systems}, IASSNS-HEP-92-1 (1992).    }

\refn\Ezawa{
Z.F. Ezawa and A. Iwazaki, {\nineit Chern-Simons Gauge Theory for
Double-layered
Electron System}, TU-402 and Nisho-18 (1992); {\nineit Quantum Hall Liquid,
Josephson Effect and Hierarchy in Double-layered Electron System}, TU-412 and
Nisho-20 (1992);
Z.F. Ezawa, A. Iwazaki and Y.S. Wu, {\nineit Josephson Effect in Multi-layered
Quantum Hall Systems}, TU-412, Nisho-21 and CTP-2168.   }

\refn\Wilczek{
F. Wilczek, \prl {69} {92} {132}.   }

\refn\Ting{
C. Ting, \ijmpB {6} {92} {3155};
{\nineit Duality in Multi-layered Quantum Hall Systems},
National Univ. of Singapore preprint NUS-HEP-92-0503 (1992).   }

\refn\DLExpt{
Y.W. Suen {\nineit et al.}, \prl {68} {92} {1379};
\semi J.P. Eisenstein {\nineit et al.}, \prl {68} {92} {1383}.   }

\refn\Halperin{
B.I. Halperin, {\nineit Helv. Phys. Acta} {\ninebf 56}, 75 (1983).   }

\refn\Hosotani{
Y. Hosotani, \prl {62} {89} {2785};  \prl {64} {90} {1691}.   }

\refn\HoHosoOne{
C.-L. Ho and Y. Hosotani, \ijmpA {7} {92} {5797}.   }

\refn\HoHosoTwo{
C.-L. Ho and Y. Hosotani, {\nineit Operator Algebra in Chern-Simons Theory on a
Torus}, TPI-MINN-92/41-T (1992)
 (To appear in {\nineit Phys. Rev. Lett.}).    }

\refn\Iengo{
R. Iengo and K. Lechner, \np {346} {90} {551}; \np {364} {91} {551};
\semi K. Lechner, \plB {273} {91} {463};
{\nineit Anyon Physics in the Torus}, Trieste SISSA preprint,  Thesis,
Apr 1991;
\semi R. Iengo, K. Lechner, and D. Li, \plB {269} {91} {109}.   }

\refn\Lee{
K. Lee, {\nineit Anyons on Spheres and Tori}, Boston Univ. report,
BU/HEP-89-28. }

\refn\Poly{
A.P. Polychronakos, \ap{203}{90}{231};
{\nineit Abelian Chern-Simons Theories and Conformal Blocks},
Univ. of Florida preprint,  UFIFT-HEP-89-9;
\plB {241} {90} {37}.     }

\refn\BraidGroup{
T. Einarsson, \prl {64} {90} {1995};  \ijmpB {5} {91} {675};
\semi  A.P. Balachandran, T. Einarsson, T.R. Govindarajan, and R. Ramachandran,
 \mplA {6} {91} {2801};
\semi T.D. Imbo and J. March-Russel, \plB {252} {90} {84};
\semi Y.S. Wu, Y. Hatsugai and M. Kohmoto, \prl {66} {91} {659};
\semi Y. Hatsugai, M. Kohmoto and Y.S. Wu, \prB {43} {91} {2661};
   {10761}.     }

\refn\WenNiu{
X.G. Wen, \prB {40} {89} {7387}; \ijmpB {4} {90} {239};
X.G. Wen and Q. Niu, \prB {41} {90} {9377}.     }


\global\newcount\secno \global\secno=0
\global\newcount\appno \global\appno=0
\global\newcount\meqno \global\meqno=1
\global\newcount\figno \global\figno=1
\newwrite\eqmac
\def\eqn#1{
        \ifnum\secno>0
            \eqno(\the\secno.\the\meqno)\xdef#1{\the\secno.\the\meqno}%
            \immediate\write\eqmac{\def\noexpand#1{\the\secno.\the\meqno}}%
        \else\ifnum\appno>0
                  \eqno(\romappno.\the\meqno)\xdef#1{\romappno.\the\meqno}%

\immediate\write\eqmac{\def\noexpand#1{\romappno.\the\meqno}}%
               \else
                  \eqno(\the\meqno)\xdef#1{\the\meqno}%
                    \fi
        \fi
        \global\advance\meqno by1
          }
\newwrite\figmac
\def\fig#1{\ifnum\secno>0
            \the\figno\xdef#1{\the\figno}%
        \fi
        \global\advance\figno by1
          }

\immediate\openout\eqmac=simons.eq

\def\firstheadline{\hfil}
\def\otherheadline{\nineit
  \ifodd\pageno \qquad \hss Multiple Chern-Simons Fields \hss Page \folio
    \else Page \folio\hss Wesolowski, Hosotani, and Ho \hss\qquad\fi}
\headline={\ifnum\pageno=1 \firstheadline \else\otherheadline\fi}



\baselineskip=9pt
\line{\ninerm Preprint from \hfil   UMN-TH-1128/93}
\line{\ninerm University of Minnesota \hfil February 25, 1993}

\vskip 2.3cm

\centerline{\bf MULTIPLE CHERN-SIMONS FIELDS ON A TORUS}

\vskip 1.5cm

\baselineskip=12pt

\centerline{\ninerm DENNE WESOLOWSKI and YUTAKA HOSOTANI}
\centerline{\eightit School of Physics and Astronomy, University of Minnesota,
  Minneapolis, MN 55455, U.S.A.}

\vskip .2cm
\centerline{\ninerm and}
\vskip .2cm

\centerline{\ninerm CHOON-LIN HO}
\centerline{\eightit Department of Physics, Tamkang University, Tamsui,
Taiwan 25137, R.O.C.}

\vskip .5cm
\centerline{\eightsl Type-set by plain \TeX }
\vskip .5cm

\baselineskip=10pt

{\eightrm
\midinsert \narrower\narrower \noindent
Intertwined multiple Chern-Simons gauge fields induce matrix statistics
among particles.  We analyse this theory on a torus, focusing on the
vacuum structure and the Hilbert space.   The theory can be mimicked, although
not completely,   by an effective theory with one Chern-Simons gauge field.
The correspondence between the Wilson line integrals, vacuum degeneracy and
wave functions for these two  theories are discussed.  Further, it
is  obtained in both of these cases that the two total momenta and
Hamiltonian commute only in the physical Hilbert space.
 \endinsert
}


\normal

\secno=1  \meqno=1

\newsection{1. Introduction \hfil}
It is known that Chern-Simons gauge theory coupled to non-relativistic matter
fields is equivalent to anyon quantum mechanics on a plane and
any Riemann surface.
The widespread interest lately in anyon quantum mechanics is due
primarily to its many applications, for instance, to the fractional quantum
Hall
effect and possibly high-$T_c$ superconductivity, but also due to the rich and
interesting mathematical properties it posesses \myref{\Reviews}.

There is growing interest in a theory in which not just one kind, but
multiple kinds of  Chern-Simons gauge interactions are introduced among
particles \myref{\WenZeeOne --\BlokWen}.
It has been known that multiple Chern-Simons interactions induce
matrix statistics which generalizes ordinary fractional statistics in the space
of particle species.

Interest  has been renewed recently \myref{\WenZeeTwo --\Ting}
by a possible application of the theory  to double-layered Hall systems
for which new experiments are now available \myref{\DLExpt}.    It also has
been
noticed that there is a connection with the Halperin wave
functions \myref{\Halperin}.

In another direction of developments,
it has  been noted that in Chern-Simons gauge theory on multiply-connected
spaces, non-integrable phases of Wilson line integrals along non-contractible
loops become physical degrees of freedom \myref{\Hosotani --\Poly}.
The dynamics of these phases lead to rich physical consequences, generating an
interesting degenerate vacuum structure.   The theory \myref{\HoHosoOne}
naturally leads to
a representation of the Braid group on multiply connected
spaces \myref{\BraidGroup}.
Also, on $T^2$ for example, the analysis can be mathematically rigorous, with
none of the infrared divergence or ambiguity in boundary conditions at space
infinity which so plague the analysis of the theory on a plane.

In this paper we shall examine multiple Chern-Simons theory on a torus.
We expect that the
study of multiple  Chern-Simons fields should shed light on the physics of
multiple-layered Hall systems, and have relevance to other condensed-matter
situations.   Furthermore we shall see that the theory has a rather beautiful
mathematical structure which by itself deserves special attention.

In Section 2 we consider the coupling of multiple ($M$) Chern-Simons
fields to matter fields, finding the statistics phases generated therefrom. We
give several examples for the case $M=2$, known to be relevant to
condensed-matter systems. In Sections 3 and 4 we turn to $T^2$ and investigate
the topologically-induced vacuum structure, deriving expressions for a
general wave function and basis of vacuum states. We then examine the result of
the action of the Wilson line operators on this basis, and consider several
examples. In Section 5 we derive results for the wave functions and action of
Wilson line integrals in the effective theory  of $M$ Chern-Simons fields. In
Section 6 we compare and make the connection between the original and
effective theories, finding their results to be almost (but not quite)
equivalent.  In Section 7 we show that the Hamiltonian and total momenta
commute among themselves only in the physical Hilbert space.
Section 8 consists of some
concluding remarks and comments.

\secno=2 \meqno=1
\newsection{2. Matrix statistics}
We consider theories with  $M$ distinct Chern-Simons gauge
fields, $a_\mu^I$ ($I=1,\cdots,M$) and nonrelativistic matter fields.
 The Lagrangian is given by $\L = \L_\CS + \L_{\rm matter}$.
The Chern-Simons term $\L_\CS$ is given by
$$\eqalign{
 \L_\CS &= {1\over4\pi} \sum_{IJ} K_{IJ} ~ a^I \epdel a^J
   \hskip .6cm (I,J=1, \cdots, M)~~,  \cr
 &(\epdel a)^\mu \equiv \eps \d_\nu a_\rho ~~~, \cr}
   \eqn\CSterm $$
where $K_{IJ}$ is a $M$-by-$M$ real symmetric matrix.

The meaning of the $K$-matrix becomes definite only when the
matter coupling is specified.  We shall consider two typical couplings.
Firstly we suppose that there are $M$ species of
matter fields $\psi_I(x)$ ($I=1,\cdots,M$), and that $\psi_I$ couples only to
$a^I_\mu$:
$$\eqalign{
\L_1= \sum_{I=1}^M & \bigg\{ i \psi_I^\dagger D_0^I \psi_I -
{1\over 2m_I} \, (D_k^I \psi_I)^\dagger (D_k^I \psi_I) \bigg\}, \cr
&D_\mu^I = \d_\mu + i a_\mu^I ~~.  \cr}
  \eqn\firstMatter $$
Secondly we suppose that there is only one  matter
field $\psi(x)$, and that $\psi$ couples to all the $a^I_\mu$'s:
$$\eqalign{
\L_2 = & ~i \psi^\dagger D_0 \psi -
{1\over 2m} \, (D_k \psi)^\dagger (D_k \psi), \cr
&D_\mu = \d_\mu + i \sum_{I=1}^M a_\mu^I ~~. \cr}
  \eqn\secondMatter $$

We shall see shortly that $\psi$ in (\secondMatter) can be viewed as a bound
state of $\psi_1$, $\cdots$, $\psi_M$ in (\firstMatter).
As is obvious, only the combination $a_\mu = \sum_{I=1}^M a_\mu^I$ is
relevant in the coupling (\secondMatter).  The effective theory for
$a_\mu$ and $\psi$ is analysed in Section 5.

To clarify the meaning of the matter couplings above, we first diagonalize the
$K$-matrix appearing in the Lagrangian. Thus
$$\eqalign{
K \, \vv^I &= \lambda_I \, \vv^I ~~~,~~~
\vv^I \cdot \vv^J = \delta_{IJ} \hskip .6cm (I,J=1, \cdots, M)~, \cr
K&= O^t \, \Lambda \, O~, \cr
O^t &= \Bigg( \vv^1, \cdots, \vv^M \Bigg) ~~~,~~~
  \Lambda = \left( \matrix{ \lambda_1 &&\cr
  &\ddots&\cr
  && \lambda_M\cr} \right)~. \cr} \eqn\diagonalizeK $$
Note that
$$O_{IJ} = (\vv^I)_J = v^I_J ~~,~~ v^I_L v^J_L = v^L_I v^L_J = \delta_{IJ}
{}~~.$$
Correspondingly, we introduce a new basis for the Chern-Simons fields:
$$ \alpha^I_\mu = O_{IJ} \, a^J_\mu ~~~,~~~ a^I_\mu = O_{JI} \, \alpha^J_\mu
  ~~~.   \eqn\newBasis $$
Then the Chern-Simons Lagrangian (\CSterm) becomes
$$\L_\CS = \sum_{I=1}^M {\lambda_I \over 4\pi}~ \alpha^I \epdel \alpha^I ~~.
          \eqn\newCSterm $$
Without loss of generality we suppose that $det\, K \not= 0$ and $\lambda_I
\not=0$.

In the  first coupling $\L_1$ we denote by $\theta^{(IJ)}_s$ the
statistics  phase  acquired by the wave function when particles of the $I$-th
and $J$-th kinds are interchanged.  (For $I\not= J$ the phase acquired
by $2\pi$-rotation is defined to be $2\theta_s^{(IJ)}$.)
Noting that $a^I= v^L_I \alpha^L$, one finds
$$\theta^{(IJ)}_s = \sum_L {\pi\over \lambda_L}
  \cdot v^L_I \cdot v^L_J = \pi ~ K^{-1}_{IJ} ~~.   \eqn\matrixStatistics $$
In other words, multiple Chern-Simons gauge theory induces matrix statistics
among particles \myref{\WenZeeOne}.
(In some of the recent literature $\theta_s^{(IJ)}$ is called a mutual or
relative statistics phase \myref{\Ezawa-\Ting}).

In the second coupling $\L_2$ one has $\sum_I a^I= \sum_I v^L_I  \, \alpha^L$.
Hence the phase $\theta_s$ picked up by interchanging two particles is
$$\theta_s = \sum_L {\pi\over \lambda_L} \cdot \Big(\sum_I v^L_I \Big)^2
 =\pi  \sum_{IJ} K^{-1}_{IJ} \equiv {\pi \over \N_{\rm eff}} ~~.
       \eqn\effectiveStatisitcs  $$
$\N_\eff$ turns out to be the Chern-Simons
coefficient in the effective theory for $a_\mu$ and $\psi$.
In the application to multi-layer Hall systems, it is related to the total
filling factor $\nu$  by  $\nu = \N_\eff^{-1}$.

Now consider  a bound state $B$ composed of $M$
particles, $\psi_1$, $\cdots$, $\psi_M$ in the first coupling $\L_1$.
If two $B$'s are interchanged,
then the statistics phase acquired is, from the preceding relations,
$$\sum_{I,J} \theta^{(IJ)}_s = \theta_s ~~~.
      \eqn\boundstateStatistics   $$
So as advertised above, we see that the second coupling may be viewed
as the effective theory  for boundstates $B$ in the first coupling.

We next consider several illustrative and relevant examples, for the
case $M=2$:

\noindent $\underline{Example ~1}$
$$\eqalign{
\noalign{\kern 2pt}
K &= \left( \matrix{ p&q\cr q&p\cr} \right) ~~~, \cr
\noalign{\kern 8pt}
\Lambda &= \left( \matrix {p+q&\cr &p-q\cr} \right) ~~,~~
O= {1\over \sqrt{2} } \left( \matrix{1&1\cr 1&-1\cr} \right)~, \cr
\noalign{\kern 8pt}
\theta^{(11)}_s &= \theta^{(22)}_s = {\pi p\over p^2-q^2} ~~,~~
\theta^{(12)}_s = -{\pi q\over p^2-q^2} \cr
\noalign{\kern 8pt}
\theta_s &= {2\pi \over p+q}~~~.  \cr
\noalign{\kern 8pt}
}      \eqn\exampleOne  $$
For specific values of $p$ and $q$ we find the following.

\noindent  $\underline{Example ~2}
$$$\eqalign{
\noalign{\kern 2pt}
(p,q) &=(3, 2)   \cr
\noalign{\kern 8pt}
\theta^{(11)}_s &= \theta^{(22)}_s = {3\pi\over 5} \next
\theta^{(12)}_s = -{2\pi\over 5} \cr
\noalign{\kern 8pt}
\theta_s &= {2\pi \over 5}~~~. \cr
\noalign{\kern 8pt}
}   \eqn\exampleTwo  $$
This case has been discussed as an alternative way of
describing the first daughter state in the fractional quantum Hall
effect \myref{\BlokWen, \WenZeeOne}.

\noindent  $\underline{Example ~3}
$$$\eqalign{
\noalign{\kern 2pt}
(p ,q) &=(3 ,1)   \cr
\noalign{\kern 8pt}
\theta^{(11)}_s &= \theta^{(22)}_s = {3\pi\over 8} \next
\theta^{(12)}_s = -{\pi\over 8} \cr
\noalign{\kern 8pt}
\theta_s &= {\pi \over 2}~~~. \cr
\noalign{\kern 8pt}
}       \eqn\exampleThree   $$
It has been argued that  this corresponds to the structure found in
double-layered Hall systems at a half filling
$\nu={1\over 2}$ \myref{\WenZeeTwo --\Ting}.

\noindent  $\underline{Example ~4}$
$$\eqalign{
\noalign{\kern 2pt}
(p,q) &=(0, 2) \cr
\noalign{\kern 8pt}
\theta^{(11)}_s &= \theta^{(22)}_s = 0 \next
\theta^{(12)}_s = {\pi\over 2} \cr
\noalign{\kern 8pt}
\theta_s &= \pi~~~ . \cr
\noalign{\kern 8pt}
}       \eqn\exampleFour   $$
Wilczek has proposed this case for the anyon superconductivity with P and T
invariance \myref{\Wilczek}.
Indeed, if one supposes that under P (or T), $\alpha_1^\mu$ and
$\alpha_2^\mu$ are interchanged in addition to the ordinary transformation,
then the theory becomes manifestly invariant.  The model should exhibit
no P or T violation  effect.

\secno=3 \meqno=1

\newsection{3. The vacuum structure on $T^2$}
On a torus, $T^2$, for each C-S field there are two nonintegrable phases of
Wilson line integrals along non-contractible loops, $C_j$.  Take a rectangular
torus with coordinates ($0 \le x_j \le L_j$, $j=1,2$).   The constant
parts of $a^I_j(x)$ are new degrees of freedom undetermined by the matter
content:
$$\exp \bigg\{ i \int_{C_j} dx_k \, a_k^I \bigg\}
\Longrightarrow W_j^I=e^{i\theta^I_j} ~~~.  \eqn\WilsonPhase  $$
The Wilson line phases $\theta^I_j$'s introduce an interesting structure
into the theory, which we are going to clarify in the following.

Inserting $a^I_j = \theta^I_j/L_j + \cdots$  into the Lagrangian $\L_\CS$,
one finds that
$$\L_\CS \Longrightarrow
 {1\over 4\pi}\,  K_{IJ} ( \theta_2^I \dot\theta_1^J -
      \theta_1^I \dot\theta_2^J ) ~~~.     $$
Hence  one sees that $\theta^I_1$'s and $\theta^I_2$'s define conjugate pairs:
$$[ \, \theta^I_1 , \theta^J_2 \,] =2\pi i \, K^{-1}_{IJ} ~~~.
       \eqn\commutatorOne $$
In the $\theta_1$-representation
$$\theta_2^I = -2\pi i \,  K^{-1}_{IJ} \, {\d\over \d \theta^J_1}
\next K_{IJ} \theta^J_2 = -2\pi i \, {\d\over \d \theta^I_1} ~~~.
     \eqn\ExpressThetaTwo $$

The system is invariant under
large gauge transformations which shift the Wilson line phases by
multiples of $2\pi$:
$$U^I_j ~:~~ \theta^I_j \go \theta^I_j + 2\pi ~~~.  \eqn\largeGTone $$
More explicitly
$$\eqalign{
a^J_\mu(x) &\go a^J_\mu(x) + \d_\mu \beta^J_{Ij}(x) ~~~,\cr
\noalign{\kern 8pt}
\beta^J_{Ij}(x) &= \delta_{IJ} ~ {2\pi x_j\over L_j} ~~~.\cr}
          \eqn\largeGTtwo   $$
Accordingly the matter field changes, in the case of the first coupling, for
instance, as
$$\psi^J(x) \go e^{-i \beta^J_{Ij}(x) } \, \psi^J(x)  ~~~.
          \eqn\largeGTthree  $$

To be precise, boundary conditions for the fields have to be specified to
define the theory on a torus.   In general the fields are not single-valued.
After translation along a noncontractible loop the fields return to their
original values up to a gauge transformation.  This problem has been analysed
in detail in the case of one Chern-Simons field in [\Hosotani, \HoHosoTwo],
and can
be straightforwardly extended to the cases under consideration.   We only
note here that transformations (\largeGTtwo) and (\largeGTthree) leave
the boundary conditions unaltered.

Unitary operators inducing the transformation (\largeGTone)
or (\largeGTtwo) are given by
$$U^I_j  =  e^{+i \ep^{jk}\, K_{IJ} \theta^J_k}    ~~~. \eqn\UnitaryIj $$
The transformation of matter fields, (\largeGTthree), is induced by
$$U^I_{{\rm matter},\, j} = \exp \bigg\{ 2\pi i \int d\x ~
{x_j\over L_j} \, \psi_I^\dagger \psi_I^{}(x) \, \bigg\} ~~~.
       \eqn\UnitaryPsiIj  $$
Since these $U^I_{{\rm matter}, \, j}$'s commute among themselves, they
will not play an important role in the following discussions.

The two sets of operators, $\{ U^I_j \}$ and $\{ W^I_j \}$, are complimentary.
They satisfy relations
$$\eqalign{
\noalign{\kern 6pt}
U^I_1 ~ U^J_2 \, &= \, e^{-2\pi i K_{IJ}} ~ U^J_2\, U^I_1 ~~~, \cr
W^I_1 \, W^J_2 &= e^{-2\pi i K^{-1}_{IJ} } \, W^J_2\, W^I_1 ~~~, \cr
U^I_j ~ W^J_k &= W^J_k ~ U^I_j ~~~. \cr
\noalign{\kern 6pt}
}  \eqn\UWalgebra  $$
Note that they do not commute with each other in general.
The algebra is invariant under the interchange of $U^I_j$ and $W^I_j$
supplemented by the replacement of $K_{IJ}$ by $K^{-1}_{IJ}$.  This suggests
that there is a duality between the theories with the Chern-Simons coefficient
matrix $K$ and with $K^{-1}$.

We would like to determine vacuum wave functions satisfying (\UWalgebra).
{}From now on we shall suppose that all $K_{IJ}$'s are integers so that all the
$U_j^I$'s commute among themselves.     We may thus simultaneously diagonalize
these operators and take
$$U^I_j \, | \Psi \ra = e^{i \gamma^I_j} \, |\Psi\ra ~~~.
           \eqn\defGamma  $$
For convenience we introduce vector notation :
$\vtheta_j=(\theta^1_j, \cdots, \theta^M_j)$,
$\vgamma_j=(\gamma^1_j, \cdots, \gamma^M_j)$, etc.
We write the wave functions
$$\eqalign{
\noalign{\kern 6pt}
u(\vtheta_1) & \equiv \la \vtheta_1 | \Psi \ra ~~~, \cr
v(\vtheta_2) & \equiv \la \vtheta_2 | \Psi \ra
  =\int d\vtheta_1 ~ \la \vtheta_2 | \vtheta_1\ra
  ~ u( \vtheta_1 ) ~~~, \cr
\noalign{\kern 6pt}
} \eqn\defWavefunction  $$
and proceed to find a general $u(\vtheta_1)$.

First consider $U_1^I |\Psi \ra = e^{i\gamma^I_1} \, |\Psi \ra $.
This implies that
$$\eqalign{
\noalign{\kern 6pt}
&e^{2\pi(\d/\d \theta^I_1)} \, u(\vtheta_1) =
  u(\cdots, \theta^I_1+2\pi, \cdots) = e^{i\gamma^I_1} \, u(\vtheta_1) ~. \cr
\noalign{\kern 6pt}
   }    $$
Thus one can express $u(\vtheta_1)$ as an
$M$-dimensional series,
$$\eqalign{
\noalign{\kern 6pt}
&u(\vtheta_1) = e^{i \vgamma_1 \vtheta_1/2\pi}
  \sum_{\vn} d(\vn) \, e^{i\vn\vtheta_1} ~~~, ~~~ \vn \in \zz^M ~~, \cr
\noalign{\kern 6pt}
}    \eqn\seriesOne  $$
where $\vn$ is an $M$-dimensional integer vector.
Secondly, $U_2^I |\Psi \ra = e^{i\gamma^I_2} \, |\Psi \ra$ gives
$$e^{-i(K \vtheta_1)^I} \, u(\vtheta_1) = e^{i\gamma^I_2} \, u(\vtheta_1) ~.
  \eqn\conditionTwo $$

We introduce here two sets of vectors by
$$\eqalign{
\noalign{\kern 6pt}
K ~ &= \Bigg( \vk_1 , \cdots , \vk_M \Bigg)
  = \left( \matrix{\vk_1^{~t} \cr \vdots \cr \vk_M^{~t}\cr} \right) ~~~, \cr
\noalign{\kern 10pt}
K^{-1} & = \Bigg( \vl_1 \, , \cdots ,  \, \vl_M \Bigg)
  = \left( \, \matrix{ \vl_1^{~t} \cr \vdots \cr \vl_M^{~t} \cr} \, \right)
   ~~~.  \cr
\noalign{\kern 6pt}
} \eqn\klVectors  $$
They satisfy
$$\vk_I \cdot \vl_J = \delta_{IJ} ~~~.  \eqn\orthogonality $$
Then we have
$$\eqalign{
e^{-i(K \vtheta_1)^I}
  \sum_{\vn} d(\vn) \, e^{i\vn\vtheta_1}
&= \sum_{\vn} d(\vn) \, e^{i(\vn - \vk_I)\vtheta_1}  \cr
&= \sum_{\vn} d(\vn + \vk_I) \, e^{i\vn \vtheta_1}  \cr}  $$
so that (\conditionTwo) leads to
$$d(\vn + \vk_I) = e^{i\gamma^I_2} \,  d(\vn) ~~~
  (I=1, \cdots, M) ~ . \eqn\conditionTwo $$

To find the general wave functions (\seriesOne) satisfying (\conditionTwo), we
first note that
$$\vgamma_2 \cdot K^{-1} \vk_I =
  \gamma_2^{J}\, (\vl_J \cdot \vk_I) = \gamma_2^I ~~.  $$
Therefore
$$\eqalign{
u(\vtheta_1) &= e^{i  \vgamma_1 \vtheta_1/2\pi}  \sum_{\vn}
  e^{i\vgamma_2 K^{-1} \vn} ~ \tild \, (\vn) ~ e^{i\vn\vtheta_1} ~~~, \cr
& \tild \, (\vn+ \vk_I) = \tild \,(\vn) ~~~.  \cr} \eqn\seriesTwo $$

Consider the $M$-dimensional lattice, $\vn \in \zz^M$. The relation
$\tild \, (\vn+ \vk_I) = \tild \,(\vn)$ in (\seriesTwo) implies that
the lattice contains $det\, K \equiv r$ inequivalent points,
each corresponding to a different ground state of the system. We
introduce $r$ linearly independent vectors corresponding to them, $\{ \vf_1,
\cdots, \vf_r \}$. We may thus parameterize $\vn$ as
$$\vn = \vf_a +
  \sum_{I=1}^M p_I \, \vk_I \qquad (a=1, \cdots, r ~~, ~~ p_I
     \in \zz)~.  \eqn\nLattice  $$
The wave function is  expressed as
$$\eqalign{
u(\vtheta_1) &= e^{i \vgamma_1 \vtheta_1/2\pi}  \sum_{a=1}^r \tild_a
  \sum_\vp e^{i(\vf_a + p_1 \vk_1 + \cdots p_M \vk_M)
  \cdot ( \vtheta_1 + K^{-1} \vgamma_2 ) } ~~~, \cr
\tild\, (\vn) &= \tild\, (\vf_a) \equiv \tild_a ~~~. \cr}
     \eqn\seriesThree $$
Since  $ \sum_I p_I \, \vk_I \cdot (\vtheta_1 + K^{-1} \vgamma_2)
 = \vp \cdot (K \vtheta_1 + \vgamma_2)$,
one then finds
$$\eqalign{
u(\vtheta_1) &= e^{i \vgamma_1 \vtheta_1/2\pi} \sum_{a=1}^r \tild_a
  \sum_\vp e^{i\vf_a \cdot (\vtheta_1 + K^{-1} \vgamma_2)} ~
  e^{i\vp \cdot (K \vtheta_1 + \vgamma_2)} \cr
&= (2\pi)^M e^{i \vgamma_1 \vtheta_1/2\pi} \sum_{a=1}^r \tild_a ~
  e^{i\vf_a \cdot (\vtheta_1 + K^{-1} \vgamma_2)} ~
  \delta_{2\pi} [ K \vtheta_1 + \vgamma_2] ~~~. \cr}
      \eqn\seriesFour $$

Let us define an equivalence relation ~$\sim$~ among vectors $ \in R^M$:
$$\vh \sim \vg \quad \Longleftrightarrow \quad
 h^I = g^I \quad (mod~ 2\pi) \quad I=1, \cdots, M ~. \eqn\equivalenceR $$
It implies, for instance, that
$\delta_{2\pi}[\,\vh\,] = \delta_{2\pi} [\,\vg\,]$  if $\vh \sim \vg$.
A set of vectors ${\cal H}(K) = \{  \vh_a \}$ , which are independent in the
coset space ~$R^M/ \sim$~ in the sense that $\vh_a \not\sim \vh_b$ iff $a\not=
b$,  is defined by
$${\cal H} (K) = \{ \, \vh_a \in R^M , ~ (a=1, \cdots, r)
  ~;~ K\, \vh_a \sim 0 \, \} ~~. \eqn\defSetH $$
With this definition one has
$$\delta_{2\pi} [~K \vphi~]
  = \sum_{b=1}^r c_b ~\delta_{2\pi} [~\phi - \vh_b~]  ~~.  \eqn\identityTwo $$

Hence our wave function is
$$\eqalign{
u(\vtheta_1) &= (2\pi)^M e^{i \vgamma_1 \vtheta_1/2\pi}
  \sum_{a=1}^r ~\tild_a ~ \sum_{b=1}^r ~ c_b ~
  e^{i\vf_a \cdot\vh_b} ~
  \delta_{2\pi} [\vtheta_1 + K^{-1} \vgamma_2 -\vh_b]  \cr
&=e^{i \vgamma_1 \vtheta_1/2\pi} \sum_{b=1}^r  g_b ~
  \delta_{2\pi} [\vtheta_1 + K^{-1}\, \vgamma_2 - \vh_b] ~~.  \cr}
    \eqn\seriesFive  $$
As an independent basis of vacua one can thus choose
$$\eqalign{
\la \vtheta_1 | \, 0_a \ra \equiv u_a(\vtheta_1)
  = e^{i \vgamma_1 \vtheta_1/2\pi}\,
  \delta_{2\pi} [\vtheta_1 + K^{-1}\, \vgamma_2 - \vh_a] ~~~,&\cr
(a=1, \cdots, ~ r= det \,K) ~~~.& \cr}  \eqn\finalWaveFunction $$
The fact that there are $r$ independent vectors $\vh_a \in {\cal H}(K)$
has been proven a posteriori.

The action of Wilson lines on the vacuum is evaluated as follows.
$$\eqalign{
\noalign{\kern 4pt}
\la \vtheta_1 \st  W_1^I \st 0_a \ra &= e^{i\theta_1^I} \, u_a(\vtheta_1)
  = e^{-i(K^{-1} \vgamma_2)^I + i h_a^I} ~ u_a(\vtheta_1) \cr
&= e^{-i \vl_I \vgamma_2 + i h_a^I} ~ u_a(\vtheta_1)  ~~~,  \cr
\la \vtheta_1 \st W_2^I \st 0_a \ra
&= e^{2\pi (K^{-1} \d/\d \vtheta_1 )^I } ~ u_a(\vtheta_1) \cr
&=u_a(\vtheta_1 + 2\pi \vl_I) ~~~. \cr
\noalign{\kern 4pt}
}  \eqn\WactionOne $$
To see that $W_2^I$ induces a mapping $I(a)$ among the vacua,
we note that
$K\, (\vh_a - 2\pi \, \vl_I) \sim 0$ so that
$\vh_a - 2\pi \vl_I \in {\cal H}$.
Hence we have
$$I(a)\, :\, \vh_a \go \vh_{I(a)}  \sim \vh_a - 2\pi \vl_I ~~~.
    \eqn\mappingI $$
Then clearly
$$ \la \vtheta_1 \st W_2^I \st 0_a \ra
  = e^{i \vl_I \vgamma_1} ~ u_{I(a)} (\vtheta_1) ~~. \eqn\WactionTwo $$
In summary,
$$\eqalign{
W_1^I ~ \st \, 0_a \ra
   &= e^{-i \vl_I \vgamma_2 - ih_a^I } ~ \st \, 0_a \ra ~~, \cr
W_2^I ~ \st \, 0_a \ra
   &= ~e^{+i \vl_I \vgamma_1} ~~ ~\st \, 0_{I(a)} \ra ~~. \cr}
   \eqn\WactionThree $$

\secno=4 \meqno=1
\newsection{4. Examples}
We apply the results in the previous section to two examples.

\vskip 8pt
\leftline{\tt Case 1. ~ $K=\left(\matrix{ 3&2\cr 2&3\cr} \right)$}

\vskip 8pt
This is the second example in Section 2.  This $K$ gives
$$K^{-1}={1\over 5} \left(\matrix{ 3&-2\cr -2&3\cr} \right) ~~~,~~~
  det\, K=5 ~~~,~~~ \N_{\rm eff} ={5\over 2} ~~~,
   \eqn\ExOneK $$
where $\N_{\rm eff}$ has been defined in (\effectiveStatisitcs).
The basis $\{ \vf_a \}$ is given by
$$\vf_a = a \, \vone \qquad (a=0, \cdots, 4)  \quad {\rm where} ~~
  \vone = \left({1\atop 1} \right) ~~~.
   \eqn\ExOneF $$
The corresponding ${\cal H} = \{ \vh_a \}$ is
$$\vh_a = {2\pi a\over 5} ~ \vone \hskip .6cm (a=0, \cdots, 4) ~.
    \eqn\ExOneH  $$

Thus the vacua are given by
$$ u_a(\vtheta_1) = e^{i \vgamma_1 \vtheta_1/2\pi} ~
  \delta_{2\pi} [~\vtheta_1 + K^{-1} \vgamma_2 - {2\pi a \over 5} ~\vone~]
  \qquad (a=0, \cdots, 4) ~.
  \eqn\ExOneU $$
They satisfy
$$\st a+5 \ra = \st a \ra ~~~.   \eqn\ExOnePeriod $$
Vectors $\vl_I$'s are given by
$$\vl_1 = {1\over 5} \left( { 3\atop -2} \right) \next
  \vl_2 = {1\over 5} \left( { -2\atop 3} \right) ~~~. $$
Since
$$\vh_a - 2\pi \vl_I \sim \vh_{a+2} ~~~,   \eqn\ExOneIa$$
one finds the mapping $I(a)$ of (\mappingI) to be
$$\st I(a) \ra = \st a+2 \ra ~~~.  \eqn\ExOneMapping  $$
Hence we have
$$\eqalign{
W_1^I ~ \st  a \ra
 &= e^{-i \vl_I \vgamma_2 + 2\pi ia/5} ~ \st  a \ra ~~, \cr
W_2^I ~ \st  a \ra
 &= ~~ e^{+i \vl_I \vgamma_1} ~~~ \st  a+2 \ra ~~. \cr}
    \eqn\ExOneW $$

\vskip 8pt
\leftline{\tt Case 2. ~ $K= \left(\matrix { 3&1\cr 1&3\cr} \right)$}

\vskip 8pt
This is the third example in Section 2.   This $K$ gives
$$K^{-1} = {1\over 8}\left(\matrix{ 3&-1\cr -1&3\cr} \right) ~~~,~~~
det\, K = 8 ~~~,~~~ \N_{\rm eff}={2} ~~~.  \eqn\ExTwoK $$
A choice of a basis for $\{ \vf_a \}$ is given by
$$\vf_a = a \, \vone - b \left( {0\atop 1} \right) \qquad (a=0 \sim
  3 ~~,~~ b=0,1)~.   \eqn\ExTwoF $$
The corresponding ${\cal H} = \{ \vh_{ab} \}$ is
$$\vh_{ab} = {\pi a\over 2} ~ \vone + {\pi b\over 4} ~ \left( {1\atop 1}
  \right) \qquad (a=0 \sim 3 ~~,~~ b=0,1)~.   \eqn\ExTwoH$$
The vacua are thus given by
$$ u_{ab}(\vtheta_1) = e^{i \vgamma_1 \vtheta_1/2\pi} ~
 \delta_{2\pi} [~\vtheta_1 + K^{-1} \vgamma_2 - \vh_{ab}~].  \eqn\ExTwoU $$
We note that
$$\eqalign{
&\st a+4,b \ra = \st a,b \ra ~~~, \cr
&\st a,b\pm 2 \ra = \st a\pm 1,b \ra ~~~. \cr}    \eqn\ExTwoPeriod$$

This time we have
$$\vl_1 = {1\over 8} \left( { 3\atop -1} \right) \next
  \vl_2 = {1\over 8} \left( { -1\atop 3} \right)
     \eqn\ExTwoL  $$
so that
$$\eqalign{
&\vh_{ab} - 2\pi \vl_1 \sim \vh_{a-1,b-1} ~~~,\cr
&\vh_{ab} - 2\pi \vl_2 \sim \vh_{a,b+1} ~~~. \cr}   \eqn\ExTwoI $$
Hence the action of Wilson line operators is given by
$$\eqalign{
W_1^I ~~~ \st \, a,b \ra
 &= e^{-i \vl_I \vgamma_2 - i\vh_{ab}} ~ \st \, a,b \ra ~~~, \cr
W_2^{(1)} ~ \st \, a,b \ra
 &= e^{+i \vl_1 \vgamma_1 } ~ \st \, a-1,b-1 \ra ~~~, \cr
W_2^{(2)} ~ \st \, a,b \ra
 &= e^{+i \vl_2 \vgamma_1 } ~ \st \, a,b+1 \ra ~~~. \cr}
     \eqn\ExTwoW  $$

\secno=5  \meqno=1
\newsection{ 5.  Effective theory}
In Section 2 we have seen that in the case of the second matter coupling
$\L_2$, (\secondMatter), the induced statistics phase $\theta_s$ is
given by (\effectiveStatisitcs).  Since the matter field $\psi$ couples
only to $a_\mu=\sum_I a_\mu^I$, one can integrate all Chern-Simons gauge fields
but the degree $a_\mu$, getting an effective theory for $a_\mu$ exactly
with  a Chern-Simons coefficient $\N_\eff$ given by (\effectiveStatisitcs).

On a torus, or more generally on a Riemann surface with a genus $\ge 1$,
however, there might arise a difference between the original multiple
Chern-Simons theory and the effective theory.  This is because the vacuum
structures are different in general.   In the next section we shall
examine the correspondence between the two theories in the cases discussed
in the previous section.

Let us recall that the effective Chern-Simons coefficient is given by
$${1\over \N_\eff} = \sum_{I,J} K^{-1}_{IJ}~~~.  \eqn\effeciveCoupling $$
Hence, even if all $K_{IJ}$'s are integers,  $\kappa_{\rm eff}$ is,
in general, a rational number,
$$\kappa_\eff ={p\over \, q\, }~~~, \eqn\rationalN $$
where $p$ and $q$ are two mutually prime integers.  In this section
we clarify the vacuum structure of this theory on a torus.
Previously it has been studied by Lee \myref{\Lee} and by
Polychronakos \myref{\Poly}.
Our derivation of  vacuum wavefunctions proceeds
along the same line as in Section 3  (We remark that our basis
is  slightly different from that given in [\Lee]  (cf. [\HoHosoTwo]).

We denote operators in the effective theory with bars on the top.
The two non-integrable phases in the theory, $\btheta_1, \btheta_2$, their
corresponding Wilson line operators, $\bW_j=e^{i\btheta_j}$, and the
generators of large gauge transformation,
$\bU_j=e^{i\epsilon^{jk}p\btheta_k/q}$
satisfy the following commutator relations:
$$\eqalign{
[\btheta_1, \btheta_2] &= {2\pi i q\over p}~~~,\cr
\bU_1 ~ \bU_2 \,&= e^{-2\pi i p/q}  ~\bU_2 \, \bU_1~~~, \cr
\bW_1 \bW_2 &= e^{-2\pi i q/p} \, \bW_2 \bW_1 ~~~.\cr}  \eqn\EffAlgebra  $$
Note that $\bU_j$ do not commute for $q\neq 1$.  But since $\bUqone$ commute
with $\bUqtwo$,  we can diagonalize $\bUqj$:
$$\bUqj \st \Psi \ra = e^{i \bgamma_j} \, \st \Psi \ra ~~.  \eqn\EffGamma $$

To find $u(\btheta_1) = \la \btheta_1 \st \Psi \ra$, we start with
$$\la \btheta_1 \st  \bUqone \st \Psi \ra
= e^{ 2\pi q (\d / \d \btheta_1) }~ u(\btheta_1)
= u(\btheta_1+ 2\pi q)
  = e^{i\bgamma_1} ~ u(\btheta_1)~~~.    \eqn\EffUone  $$
There are  $q$ independent solutions to  Eq.\ (\EffUone):
$$\eqalign{
u_s(\btheta_1) &= \sum_m c_{sm} ~e^{i ( m + ps/q
+ \bgamma_1/2\pi q ) \btheta_1 } ~~~,\cr
&\qquad\qquad s=0,1,\ldots ,q-1 ~~~.\cr} \eqn\EffSeriesOne $$

Secondly we have
$$\la \btheta_1 \st  \bUqtwo \st \Psi \ra   = e^{-ip\btheta_1}
u(\btheta_1) = e^{i\bgamma_2} ~u(\btheta_1) ~~~. \eqn\EffUtwo  $$
Substitution of (\EffSeriesOne) into (\EffUtwo) gives the condition on
$c_{s,m}$: $c_{s,m+p} = e^{i\bgamma_2} ~ c_{s,m}$,
which is solved by
$$c_{s,m} = e^{im\bgamma_2/p} ~ A_s~d_m ~~~, ~~~ d_{m+p} = d_m ~~~ .
       \eqn\EffSeriesTwo $$
The $s$-dependent constant $A_s$ will be determined later.

Now we can rewrite $u_s(\btheta_1)$ by making use of (5.7) and writing
$m=lp+r$.  Noting that $d_m=d_{lp+r}=d_r$, we have
$$\eqalign{
u_s(\btheta_1)
&= A_s \sum_{r=0}^{p-1} d_r \sum_{l=-\infty}^\infty
 \exp \bigg\{ i\Big( l+ {r\over p} \Big) \bgamma_2 +
i \Big( lp + r +
 {ps\over q}  + {\bgamma_1\over 2\pi q } \Big) \btheta_1 \bigg\} \cr
&= 2\pi A_s \sum_{r=0}^{p-1} d_r ~
  e^{ i r( \btheta_1  + \bgamma_2/p)  + i \left(ps/q + \bgamma_1/2\pi q\right)
        \btheta_1}   ~
\sum_{a=0}^{p-1} \delta_{2\pi p} [\, p\btheta_1  + \bgamma_2  - 2\pi a q\,] ~~.
        \cr}    \eqn\EffSeriesThree $$
In the second line we have expressed $\delta_{2\pi}[p\btheta_1 + \bgamma_2]$
in terms of $\delta_{2\pi p}[ \cdots ]$, taking advantage of the mutually prime
nature of $p$ and $q$.  Further
$$\eqalign{
u_s(\btheta_1)
&= 2\pi A_s ~e^{i \left(ps/q + \bgamma_1/2\pi q\right)  \btheta_1}
\sum_{r=0}^{p-1} \sum_{a=0}^{p-1} d_r  ~e^{i 2\pi rqa/p}
\delta_{2\pi p} [\, p\btheta_1 + \bgamma_2  - 2\pi a q\, ] \cr
&= 2\pi A_s~
e^{i \left(ps/q + \bgamma_1/2\pi q\right)  \btheta_1}
\sum_{a=0}^{p-1} f_a  ~ \delta_{2\pi} \Big[ \btheta_1 +
{1\over p} (\bgamma_2  - 2\pi a q) \Big]  ~~~. \cr}
  \eqn\EffSeriesFour $$
We choose $A_s$ such that $u_{s+q}(\btheta_1)=u_s (\btheta_1)$,
which leads to the condition on $A_s$:
$A_{s+q} = e^{i\bgamma_2}~A_s$.  Again, this is solved by
$A_s=e^{is\bgamma_2/q}$.
Hence as a linearly independent basis one can take
$$\eqalign{
\noalign{\kern 4pt}
u_{a,s}(\btheta_1) =  e^{i \bgamma_1 \btheta_1 /2\pi q}  ~
e^{is (\bgamma_2 + p \btheta_1) /q}  ~ \delta_{2\pi}
  \Big[ \, \btheta_1 + {1\over p} (\bgamma_2 - 2\pi aq)\, \Big] ~~~,&  \cr
(a=0, \cdots, p-1 \next s=0, \cdots, q-1 ) ~~~.& \cr
\noalign{\kern 4pt}
}   \eqn\EffBasis   $$

Let us denote the corresponding vacuum by $\st a,s \ra$.  They satisfy
$$\eqalign{
\st a+p,s \ra &= \st a,s \ra ~~~, \cr
\st a,s+q \ra &= \st a,s \ra ~~~. \cr}   \eqn\EffIdentity $$
The actions of $\bU_j$ and $\bW_j$ on these vacua are found to be :
$$\eqalign{
\bU_1 \, \st a,s \ra &= e^{i(\bgamma_1 + 2\pi sp)/q } \, \st a,s \ra ~~~,\cr
\bU_2 \, \st a,s \ra &=~  e^{i\bgamma_2 /q } \, ~ \st a,s-1 \ra ~~~,\cr
\bW_1 \, \st a,s \ra &= e^{- i(\bgamma_2 - 2\pi aq)/p } \, \st a,s \ra ~~~,\cr
\bW_2 \, \st a,s \ra &= ~e^{i\bgamma_1 /p } \, ~\st a-1,s \ra ~~~. \cr}
  \eqn\EffUWaction  $$
Notice that $\bU_j$ acts on the space of the $s$ index, whereas $\bW_j$
on that of the $a$ index.

There can be two possible interpretations of  (\EffUWaction).
One can view that $\bU_2$ generates gauge copies so that there are only
$p$ physically distinct vacuum states.   This viewpoint has been adopted
by Polychronakos \myref{\Poly}.   An alternative interpretation is possible.
In applications to condensed-matter systems, vacua
here correspond to ground states.  Wen and Niu \myref{\WenNiu}
have argued that additional
interactions such as a tunneling effect in condensed-matter systems generally
lift  the degeneracy in the $s$ index, which forces us to regard all
$p\cdot q$ states physically distinct.  At the level of an idealized
Chern-Simons
gauge theory under consideration, both interpretations are acceptable.

\secno=6 \meqno=1
\newsection{6. The correspondence between the two theories}
There is a correspondence between  multiple Chern-Simons  theory
and single Chern-Simons theory with the corresponding effective coupling.
We work out this correspondence in two typical cases discussed in Section 4.

First
$${\tt case ~1~ :} \hskip 1.5cm
K= \left(\matrix{ 3&2\cr 2&3\cr} \right)   \next
 \N_\eff= {p\over \, q\, }={5\over \, 2\, } ~~~. \eqn\KNone  $$
Recall that
$$det \, K = 5 \next p\, q= 10 ~~~. \eqn\degeneracyOne $$
Hence in the $K$-theory there are 5 distinct vacua, whereas in the effective
theory there are 10 vacua (including gauge copies in Polychronakos'
interpretation).

Applying the results in the previous section, one finds that in the effective
theory
$$\eqalign{
\noalign{\kern 4pt}
\bU_1 \, \st a,s \ra &= e^{i\bgamma_1 /2 } \, \st a,s \ra ~~~,\cr
\bU_2 \, \st a,s \ra &= e^{i\bgamma_2 /2 } \, \st a,s-1 \ra ~~~,\cr
\bW_1 \, \st a,s \ra &= e^{-i(\bgamma_2 - 4\pi a)/5 } \, \st a,s \ra ~~~,\cr
\bW_2 \, \st a,s \ra &= e^{i\bgamma_1 /5 } \, \st a-1,s \ra ~~~. \cr
\noalign{\kern 4pt}
}        \eqn\OneUWaction $$
On the other hand in the $K$-theory we have seen in Section 3 that
$$\eqalign{
\noalign{\kern 4pt}
U_j^{(1)} ~~U_j^{(2)} \st a \ra &= e^{i(\gamma_j^1+\gamma_j^2)}
  ~\st a \ra ~~~, \cr
W_1^{(1)} ~W_1^{(2)} \st a \ra
 &= e^{4\pi ia/5 - i(\vl_1 + \vl_2) \vgamma_2} ~\st a \ra \cr
 &= e^{4\pi ia/5 - i(\gamma_2^1 + \gamma_2^2)/5} ~\st a \ra ~~~, \cr
W_2^{(1)} ~W_2^{(2)} \st a \ra
 &= e^{+i(\gamma_1^1 + \gamma_1^2)/5 } ~\st a+4 \ra \cr
 &= e^{+i(\gamma_1^1 + \gamma_1^2)/5 } ~\st a-1 \ra ~~~. \cr
\noalign{\kern 4pt}
}       \eqn\OneUWactionTwo $$

Comparing (\OneUWaction) and (\OneUWactionTwo),  one finds an
exact correspondence:
{%
\def\space{height8pt &\omit  &&\omit &&\omit  &\cr}
\def\smallspace{height5pt &\omit  &&\omit &&\omit  &\cr}

\def\hh{\hskip .5cm}
\def\hs{\hskip .2cm}
$$\vcenter{ \hbox{
\vtop{\offinterlineskip
\hrule
\halign{&\vrule# &\strut\hh \hfil $\big #$\hfil \hh &#
                 &\strut\hs \hfil $\big #$\hfil \hs &#
                 &\strut\hh \hfil $\big #$\hfil \hh &#\vrule \cr
\smallspace
&K  &&: &&\N_\eff  &\cr
\smallspace
\noalign{\hrule}
\space
&\theta_j^{(1)} + \theta_j^{(2)} &&: &&\btheta_j &\cr
\space
&U_j^{(1)} U_j^{(2)} &&: &&(\bU_j)^2 &\cr
\space
&W_j^{(1)} W_j^{(2)} &&: &&\bW_j &\cr
\space
&\gamma_j^{(1)} + \gamma_j^{(2)} &&: &&\bgamma_j &\cr
\space
&\st a \ra &&: &&\st a,s \ra &\cr
\space
\noalign{\hrule}
} }     } } \eqn\tableOne $$
}
Note that both $\st a, s=0 \ra$ and $\st a, s=1 \ra$ in the effective theory
correspond to $\st a \ra$ in the multiple Chern-Simons theory.

The second case is
$${\tt case ~2~ :} \hskip 1.5cm
K= \left(\matrix{ 3&1\cr 1&3\cr} \right)   \next
 \N_\eff= {p\over \, q\, }= 2 ~~~. \eqn\KNtwo  $$
We recall that
$$det \, K = 8 \next p\, q= 2 ~~~. \eqn\degeneracyTwo $$
Hence in the $K$-theory there are 8 distinct vacua, whereas in the effective
theory there are only two vacua.  Nevertheless
there exists correspondence.

In the effective theory we have found that
$$\eqalign{
\noalign{\kern 4pt}
\bU_j \, \st c \ra &= e^{i\bgamma_j} \, \st c \ra ~~~, \cr
\bW_1 \, \st c \ra &= e^{i\pi c - i\bgamma_2/2 } \, \st c \ra ~~~, \cr
\bW_2 \, \st c \ra &= e^{i\bgamma_1/2} \, \st c-1 \ra ~~~,  \cr
\noalign{\kern 4pt}
}        \eqn\TwoUWaction $$
where $\st c+2 \ra = \st c \ra$.
In the $K$-matrix theory of Section 3
$$\eqalign{
\noalign{\kern 4pt}
U_j^{(1)} \,~U_j^{(2)} \,\st a,b \ra &= ~~e^{i(\gamma_j^1+\gamma_j^2)}
 ~~\st a,b \ra ~~~, \cr
W_1^{(1)} \, W_1^{(2)} \st a,b \ra
 &= e^{i\pi a - i\pi b/2 - i(\gamma_2^1 + \gamma_2^2)/5} ~\st a,b \ra ~~~,\cr
W_2^{(1)} \, W_2^{(2)} \st a,b \ra
 &= ~ e^{+i(\gamma_1^1 + \gamma_1^2)/4 } \, \st a-1,b \ra ~~~,\cr
\noalign{\kern 4pt}
}       \eqn\TwoUWactionTwo $$
with the identities (\ExTwoPeriod).

The correspondence is found to be
{%
\def\space{height8pt &\omit  &&\omit &&\omit  &\cr}
\def\smallspace{height5pt &\omit  &&\omit &&\omit  &\cr}

\def\hh{\hskip .5cm}
\def\hs{\hskip .2cm}
\def\onehalf{ {\hbox{${1\over 2}$}} }
$$\vcenter{ \hbox{
\vtop{\offinterlineskip
\hrule
\halign{&\vrule# &\strut\hh \hfil $\big #$\hfil \hh &#
                 &\strut\hs \hfil $\big #$\hfil \hs &#
                 &\strut\hh \hfil $\big #$\hfil \hh &#\vrule \cr
\smallspace
&K  &&: &&\N_\eff  &\cr
\smallspace
\noalign{\hrule}
\space
&\theta_j^{(1)} + \theta_j^{(2)} &&: &&\btheta_j &\cr
\space
&U_j^{(1)} U_j^{(2)} &&: &&(\bU_j)^2 &\cr
\space
&W_j^{(1)} W_j^{(2)} &&: &&\bW_j &\cr
\space
&\onehalf \big( \gamma_1^{(1)} + \gamma_1^{(2)} \big)  &&: &&\bgamma_1 &\cr
\space
&\onehalf \big( \gamma_2^{(1)} + \gamma_2^{(2)} \big)  + \pi b
                    &&: &&\bgamma_2 &\cr
\space
&\bigg\{ \matrix{\st c, b \ra \cr \st c+2, b \ra \cr}  &&: &&\st c \ra &\cr
\space
\noalign{\hrule}
} }     } } \eqn\tableTwo $$
}

Note that for each $b$ (0 or 1) the four states $\st a,b \ra$ ($a=0 \sim 3$)
close under the
operations above. Also, $\st a,b \ra$ and $\st a+2,b \ra$ in the $K$-theory
correspond to the same state in the effective theory.  Indeed, in terms of
$\theta_1^{(1)} \pm \theta_1^{(2)} \equiv \theta_1^{\pm}$~,  the vacuum
wave function in the $K$-theory yields
$$u_{ab} (\vtheta_1) \rightarrow e^{i(\gamma_1^{(1)} + \gamma_1^{(2)})
  \theta_1^+ /4\pi} ~~
 \delta_{2\pi} \Big[ \,\theta_1^+ + {1\over 4}(\gamma_2^{(1)}
  +\gamma_2^{(2)}) - \pi a + {\pi b \over 2}\, \Big] ~~~.
      \eqn\TwoWaveFunction$$
Both $u_{a,b}(\vtheta_1)$ and $u_{a+2,b}(\vtheta_1)$ give the same $\theta_1^+$
wave function.  The $\theta_1^-$ parts  are different,
but the effective theory is blind to them.

\secno=7  \meqno=1

\newsection{7.  The Hilbert space}
It has been shown that the  Hamiltonian and two total momenta
in  theories defined on a torus with a single Chern-Simons field
 commute among themselves only in the
physical Hilbert space \myref{\HoHosoTwo}.  In this section we examine
the Hilbert space
of multiple Chern-Simons theory with two kinds of matter couplings,
$\L_1$ and $\L_2$ defined in Section 2.

The argument proceeds as in ref.\ [\HoHosoTwo].   We first solve field
equations to express Chern-Simons fields in terms of Wilson line phases
and matter fields.   Then we examine commutators among the Hamiltonian and
momenta.

First consider the first matter coupling $\L_1$, (\firstMatter).  Chern-Simons
field equations are
$${1\over 4\pi} \, K_{IJ} \, \eps f_{\mu\nu}^J = j^\mu_I ~~~,
   \eqn\OneCSequation $$
where $j^\mu_I$ is a current of the $I$-th matter field $\psi_I$.  Solving
them in the radiation gauge, one finds
$$\eqalign{
\noalign{\kern 4pt}
a^j_I(x) &= {\theta^I_j(t) \over L_j}
   - {\Phi_I\over 2L_1L_2} \, \ep^{jk} x_k + \hat a^j_I(x) ~~~, \cr
\hat a^j_I(x) &= \int d\y ~ \ep^{jk} \nabla^x_k G(\x-\y) ~
  \bigg\{ 2\pi K^{-1}_{IJ} \, \psi^\dagger_{J} \psi^{}_{J}(y)
      + {\Phi_I\over L_1L_2}   \bigg\}   ~~~. \cr
\noalign{\kern 4pt}
}          \eqn\OneExpressCS $$
Here $\Phi_I= - \int d\x \, f^I_{12}$
and $G(\x)$ is the periodic Green's function on a torus satisfying
$\Delta \, G(\x) =  \delta(\x) - (1/L_1L_2)$.

The field equations (\OneCSequation) are solved by (\OneExpressCS) except
for
$$K^{-1}_{IJ} \, Q_J^{} + {\Phi_I\over 2\pi} \approx 0  \hskip 1.5cm
    (I=1, \cdots, M)
   \eqn\OneConstraint $$
where $Q_I= \int d\x \, j^0_I$.  The conditions (\OneConstraint) have to be
imposed as constraints on physical Hilbert states.
(We have adopted the notation
$\approx$ to indicate this.)  We note that $Q_I$ is conserved
thanks to the gauge invariance, or the field equation for $\psi_I$, whereas
$\Phi_I$ is a constant fixed by boundary conditions for $a^I_\mu$.

The Hamiltonian and momenta are given by
$$\eqalign{
\noalign{\kern 4pt}
H &= \int d\x ~ \sum_I {1\over 2m_I} \,
       (D_k^I \psi_I)^\dagger (D_k^I \psi_I) ~~~, \cr
P^j &= -i \int d\x ~ \sum_I \psi_I^\dagger D_j^I \psi_I^{}
 =  -i \int d\x ~ \sum_I \psi_I^\dagger \DbarjI \psi_I^{} ~~~,  \cr
\DbarjI &= \nabla_j - i {\theta_j^I\over L_j}
   + i {\Phi_I\over 2L_1L_2} \, \ep^{jk}x_k  ~~~. \cr
\noalign{\kern 4pt}
}           \eqn\OneHP  $$
In  $D_j^I \psi_I$ in the expression above,
 (\OneExpressCS) is to be substituted .  Note that the
Hamiltonian is not completely normal-ordered.  The ordering of $\psi_I$
operators must be taken as it appears.

Useful relations are
$$\eqalign{
\noalign{\kern 4pt}
[P^k , \, \theta^I_j\,]
  &= - 2\pi i \, \ep^{kj} {1\over L_k} \, K^{-1}_{IN}\, Q_{N}^{} ~~~, \cr
[\,H\, , \, \theta^I_j\, ]
  &= - 2\pi i \, \ep^{kj} {1\over L_k} \, K^{-1}_{IN}\, J^k_{N} ~~~, \cr
\noalign{\kern 4pt}
}       \eqn\usefulOne $$
where $J^k_I = \int d\x \, j^k_I(x)$.  Further
$$\eqalign{
[P^j , \psi_I ] &= i \DbarjI \psi ~~, \cr
\noalign{\kern 8pt}
\bigg( \matrix{ [P^j, D^I_k \psi_I ] \cr
         [P^j, \DbarkI \psi_I ] \cr}  \bigg)
 & = \bigg( \matrix{ i\DbarjI  D^I_k \psi_I \cr
                      i\DbarjI  \DbarkI \psi_I \cr} \bigg)
 - \ep^{jk} \,{\Phi_I\over L_1L_2} \, \psi_I
-\ep^{jk} {2\pi\over L_1L_2} \, K^{-1}_{IJ} \psi_I Q_J ~~. \cr}
      \eqn\usefulTwo  $$

With the aid of (\usefulOne) and (\usefulTwo) it is straightforward to find
$$\eqalign{
[P^j,P^k] &= i \ep^{jk} \, {2\pi\over L_1L_2}
   ~ Q_I \, \bigg( K^{-1}_{IN} Q_{N}^{} + {\Phi_I \over 2\pi} \bigg) ~~~, \cr
\noalign{\kern 5pt}
[P^j,\, H\, ] &= i \ep^{jk} \, {2\pi\over L_1L_2}
   ~ J^k_I \, \bigg( K^{-1}_{IN} Q_{N}^{} + {\Phi_I \over 2\pi} \bigg) ~~~.
\cr}
    \eqn\OnePHcommutator  $$
Here we have made use of the fact that gauge invariant quantities
are single-valued on a torus.    We observe that $H$ and $P^j$ commute
among themselves only in the physical Hilbert space defined by
(\OneConstraint).

Similarly commutators between $W^I_j$ and $(H,P^j)$ are evaluated.  We note
that
$$\eqalign{
[W^I_j, Q_N] &=0  ~~~, \cr
[W^I_j, J^k_N\,] &= \ep^{jk} {2\pi\over m_N L_k} \, K^{-1}_{IN} \, Q_N^{}
  W^I_j   ~~~.  \cr}    \eqn\OneWJ  $$
We find that
$$\eqalign{
[W^I_j, P^k] &= \ep^{jk} {\pi\over L_k} \, K^{-1}_{IN}\,
         \{ Q_N , W^I_j \} ~~~, \cr
\noalign{\kern 5pt}
[W^I_j, \,H\,] &= \ep^{jk} {\pi\over L_k} \, K^{-1}_{IN}\,
         \{ J^k_N \, , W^I_j \} ~~~. \cr}   \eqn\OneWPHcommutator $$

Next we consider the second matter coupling $\L_2$, (\secondMatter).
Field equations are
$${1\over 4\pi} \, K_{IJ} \, \eps f_{\mu\nu}^J = j^\mu ~~~.
   \eqn\TwoCSequation $$
All Chern-Simons fields couple to the same current $j^\mu$.
Inverting (\TwoCSequation), one finds that the $I$-th Chern-Simons field
$a_\mu^I$ has a coupling strength $\N_I$ to $\psi$:
$$ {1\over \N_I} = \sum_J K^{-1}_{IJ}   \next
{1\over \N_\eff} = \sum_I {1\over \N_I} ~~~.   \eqn\effICScoupling $$
Hence
$$\eqalign{
a^j_I(x) &= {\theta^I_j(t) \over L_j}
   - {\Phi_I\over 2L_1L_2} \, \ep^{jk} x_k + \hat a^j_I(x) ~~~, \cr
\hat a^j_I(x) &= \int d\y ~ \ep^{jk} \nabla^x_k G(\x-\y) ~
  \bigg\{ {2\pi\over \N_I}\, \psi^\dagger \psi(y)
      + {\Phi_I\over L_1L_2}   \bigg\}   ~~~. \cr}
        \eqn\TwoExpressCS $$

There still are $M$ constraints:
$${Q\over \N_I} + {\Phi_I\over 2\pi} \approx 0  \hskip 1.cm
    (I=1, \cdots, M) ~~.
   \eqn\TwoConstraint $$
Summing over $I$ in the above, one has
$${Q\over \N_\eff} + {\Phi_{\rm tot}\over 2\pi} \approx 0  \next
    \Phi_{\rm tot} = \sum_I \Phi_I ~~~.
   \eqn\TwoConstraintTotal $$

The Hamiltonian and momenta are given by
$$\eqalign{
H &= \int d\x ~  {1\over 2m} \,
       (D_k \psi)^\dagger (D_k \psi) ~~~, \cr
P^j &=  -i \int d\x ~  \psi^\dagger \Dbar_j \psi ~~~,  \cr
\Dbar_j &= \nabla_j - i  \sum_I {\theta_j^I \over L_j}
   + i {\Phi_\tot \over 2L_1L_2} \, \ep^{jk}x_k  ~~~. \cr}
         \eqn\TwoHP  $$
Relations corresponding to (\usefulOne) and (\usefulTwo) are
$$\eqalign{
[P^k , \theta^I_j]
  &= - 2\pi i \, \ep^{kj} {1\over L_k} \,{1\over \N_I} Q ~~,\cr
[\,H\, , \theta^I_j]
  &= - 2\pi i \, \ep^{kj} {1\over L_k} \,{1\over \N_I} J^k ~~,\cr
[P^k , \psi ] &= i \Dbar_k \psi ~~, \cr
\noalign{\kern 6pt}
\bigg( \matrix{ [P^j, D_k \psi] \cr [P^j, \Dbar_k \psi] \cr} \bigg)
 &= \bigg( \matrix{ i\Dbar_j  D_k \psi\cr
                     i\Dbar_j  \Dbar_k \psi\cr} \bigg)
 - \ep^{jk} \,{\Phi_\tot \over L_1L_2} \, \psi
-\ep^{jk} {2\pi\over L_1L_2} \,{1\over \N_\eff}\, \psi Q ~~. \cr}
      \eqn\usefulThree  $$
It is easy to see now that
$$\eqalign{
[P^j,P^k] &= i \ep^{jk} \, {2\pi\over L_1L_2}
   ~ Q \, \bigg( {Q\over \N_\eff} + {\Phi_\tot \over 2\pi} \bigg) ~~~, \cr
\noalign{\kern 5pt}
[P^j,\, H\, ] &= i \ep^{jk} \, {2\pi\over L_1L_2}
   ~ J^k \, \bigg( {Q\over \N_\eff} + {\Phi_\tot \over 2\pi} \bigg) ~~~. \cr}
    \eqn\TwoPHcommutator $$
Further
$$\eqalign{
[W^I_j, Q_N] &=0  ~~~, \cr
[W^I_j, J^k_N\,] &= \ep^{jk} {2\pi\over m_N L_k} \, K^{-1}_{IN} \, Q_N^{}
  W^I_j   ~~~,  \cr}    \eqn\TwoWJ  $$
and
$$\eqalign{
[W^I_j, P^k] &= \ep^{jk} {\pi\over L_k} \,{1\over \N_I} \,
         \{ Q\, , W^I_j \} ~~~, \cr
\noalign{\kern 5pt}
[W^I_j, \,H\,] &= \ep^{jk} {\pi\over L_k} \,{1\over \N_I} \,
         \{ J^k  , W^I_j \} ~~~. \cr}   \eqn\TwoWPHcommutator $$
Note that (\TwoPHcommutator) and (\TwoWPHcommutator) are formally obtained
from (\OnePHcommutator) and (\OneWPHcommutator) by identifying
$Q_I=Q$ and $J^k_I=J^k$.

Finally we define
$$ \bW_j = \prod_I W^I_j  ~~~.   \eqn\newDefWbar $$
This corresponds to the Wilson line operator in the effective theory discussed
in Section 5.  In the second matter coupling $\L_2$, commutators between
$\bW_j$ and $(H,P^k)$ are evaluated from (\TwoWPHcommutator).  For $H$, one has
$$\eqalign{
[\bW_j, \,H\, ] &= \ep^{jk} {\pi\over L_k}
 \sum_I {1\over \N_I} \, W^1_j \cdots W^{I-1}_j ( W^I_j  J^k + J^k W^I_j)
    W^{I+1}_j \cdots W^M_j \cr
&=  \ep^{jk} {\pi\over L_k} \sum_I {1\over \N_I} \bigg\{ ~
\bW_j J^k - \ep^{jk} {2\pi\over mL_k} \sum_{N=I+1}^M {1\over \N_N}\, Q\bW_j \cr
&\hskip 2.4cm + J^k \bW_j +  \ep^{jk} {2\pi\over mL_k}
    ~\, \sum_{N=1}^{I-1}~\, {1\over \N_N}\, Q\bW_j ~ \bigg\} \cr
&=\ep^{jk} {\pi\over L_k} \, {1\over \N_\eff} \, \{ \bW_j , J^k \}  ~~~. \cr}
    \eqn\TwoWbarHcommutator $$
In the second equality we have made use of (\TwoWJ).  The evaluation
of the commutator $[\bW_j,P^k]$ is simpler as $W^I_j$ and $Q$ commute.
To summarize one finds
$$\eqalign{
[\bW_j, P^k] &= \ep^{jk} {\pi\over L_k} {1\over \N_\eff} \,
      \{ \,Q\, ,\bW_j \} ~~~, \cr
[\bW_j, \, H\, ]&= \ep^{jk} {\pi\over L_k} {1\over \N_\eff} \,
      \{ J^k ,\bW_j \} ~~~. \cr}
    \eqn\WbarPHcommutator $$

The algebra defined by (\TwoPHcommutator) and (\WbarPHcommutator) is exactly
the algebra of the effective theory which has been previously obtained in
ref. [\HoHosoTwo].  We also remark that the same commutator relations among
$H$,
$P^k$,  and $W^I_j$ hold  in relativistic Dirac theories, as well.

\secno=8  \meqno=1

\newsection{8. Conclusion}
In this paper we have unveiled some of the rich mathematical
structure of Chern-Simons gauge theory on a torus.
In addition to inducing matrix statistics among particles, the theory of
multiple Chern-Simons gauge fields enriches the vacuum structure.
Depending on the Chern-Simons coefficient matrix $K$,
sometimes the theory can be mapped to an effective theory with just one
Chern-Simons gauge field, but in general it contains greater degeneracy in
the vacua.  We have worked out two typical examples, finding rather interesting
correspondence between the two theories.

Also we have examined the algebra generated by the Hamiltonian, momenta, and
Wilson line operators in the multiple Chern-Simons theory.  We have found that
the  Hamiltonian and momenta commute among themselves only in the physical
Hilbert space.

Having clarified the vacuum structure of multiple Chern-Simons theory, one
might ask what the Schr\"odinger equation for many particle systems is and
whether or not there exists a singular gauge transformation which eliminates
all interactions among particles except for the effect of
the matrix statistics.  Further, the multiple Chern-Simons theory must
lead to representations of a generalized Braid group algebra.
These problems are reserved for future investigation.

\newsection{Acknowledgements}
This work is supported in part by the U.S.\ Department of Energy under
Contract No.\ DE-AC02-83ER-40105 (D.W.\ and Y.H.), and by R.O.C.\ Grant
NSC 82-0208-M032-020 (C.-L.H.).  D.W. would like to thank
K.\ Middleton-Tondra and C.A.\ Jones for stimulating discussions.
C.-L.H. would like to thank the staff of the High Energy Theory Group and
TPI at the University of Minnesota for their hospitality and financial support.

\newsection{References}

\ninerm
\baselineskip=12pt plus 1pt minus 1pt
 \immediate\closeout\reffile
	\input refs.tmp\vfill\eject\nonfrenchspacing

\vfil
\bye